\documentclass{sig-alternate-10pt}

\usepackage[utf8]{inputenc}
\usepackage{authblk}
\usepackage{enumitem}
\usepackage{float}
\usepackage{booktabs}
\usepackage{float}
\usepackage{subcaption}
\usepackage[hyphens]{url}
\usepackage[hidelinks]{hyperref}
\hypersetup{breaklinks=true}
\graphicspath{{figures/}}

\usepackage{lmodern}
\begin{document}
\sloppy
	
\title{The Challenges of Trace-Driven Wi-Fi Emulation}

\author[$\dag$,$\P$,$\S$]{Mohammad Imran Syed}
\author[$\dag$]{Renata Teixeira}
\author[$\dag$]{Sara Ayoubi}
\author[$\dag$]{Giulio Grassi}
\affil[$\dag$]{INRIA, Paris}
\affil[$\P$]{Sorbonne Sciences University, Paris}
\affil[$\S$]{EIT Digital Master School}
\affil[ ]{\textit {\{mohammad.syed,  renata.teixeira, sara.ayoubi, giulio.grassi\}@inria.fr}}

\maketitle

\begin{abstract}
Wi-Fi link is unpredictable and it has never been easy to measure it perfectly; there is always bound to be some bias. As wireless becomes the medium of choice, it is useful to capture Wi-Fi traces in order to evaluate, tune, and adapt the different applications and protocols. Several methods have been used for the purpose of experimenting with different wireless conditions: simulation, experimentation, and trace-driven emulation. In this paper, we argue that trace-driven emulation is the most favorable approach. In the absence of a trace-driven emulation tool for Wi-Fi, we evaluate the state-of-the-art trace-driven emulation tool for Cellular networks and we identify issues for Wi-Fi: interference with concurrent traffic, interference with its own traffic if measurements are done on both uplink and downlink simultaneously, and packet loss. We provide a solid argument as to why this tool falls short of effectively capturing Wi-Fi traces. The outcome of our analysis guides us to propose a number of suggestions on how the existing tool can be tweaked to accurately capture Wi-Fi traces.
\end{abstract}




\section{Introduction}
\label{sec:intro}
\hfill\

Mobile networks are becoming increasingly more popular than wired networks due to the widespread use of mobile devices (e.g. smartphones, laptops, tablets, smartwatches, etc.). The number of smartphone users alone is expected to reach 2.87 billion by 2020 \cite{smartphone-users}. The portability, availability, affordability, and increasing speeds of wireless connections, have made wireless the medium of choice.

The quality of wireless connectivity varies drastically from place to place depending on the coverage. There are a number of factors that affect signal quality or create interference, like poor network configuration, old equipment, fluctuating demands of users, router position, congestion, and coverage. As many of today's applications and services will be running over Wi-Fi or Cellular, it is useful to evaluate the performance of these applications in different wireless networks. For instance, application developers may wish to understand the impact of Wi-Fi packet-drop on their application, or what are the user-perceived latencies over Wi-Fi versus LTE.

There are different options for evaluating the applications and services in real network environments, namely simulation, experimentation, and emulation. \textbf{Simulation} is the easiest way to experiment with different wireless network conditions. Simulators are used to mimic the behavior of a certain network in a software-based environment. They offer different topologies, different network entities like routers, nodes, access points, etc., and tuning of real network parameters like packet loss, jitter, delay, and latency. There exist a number of wireless simulation tools \cite{ns2}, \cite{ns3}, \cite{omnet++}, \cite{opnet}, \cite{netsim}, \cite{Qualnet:2014}, \cite{7561126}. Repeatability, control, configurability, and experiments of large-scale networks are the advantages of simulation. The main limitation of simulation tools, however, is that they require the user to tune different parameters e.g., level of interference, congestion, loss rate, etc. which may not reflect real wireless network conditions.

At the other end of the spectrum, there is \textbf{Experimentation}, where developers evaluate their applications over deployed wireless links, either over testbeds or by relying on volunteer testers. The results of such experiments capture the impact of real wireless network conditions. One disadvantage of experimentation is that it offers no repeatability. The variability of wireless networks makes it hard to reproduce the results. The results of experimentation are, therefore, hard to interpret and one can not distinguish the issues with application versus wireless issues.



Finally, \textbf{trace-driven emulation} involves recording real wireless traces and later replaying them under emulated network conditions. The clear benefit of trace-driven emulation is its ability to capture real network conditions and the repeatability of the experiments. One can run the same network conditions several times, which eases application or system debugging, and enables comparative analysis of different applications or protocols over the same network conditions.


While there exist trace-driven emulation tools for cellular \cite{180329} and web traffic \cite{Netravali:2015:MAR:2813767.2813798}, to the best of our knowledge, there exists no such tool for Wi-Fi. In this paper, we evaluate how well the state-of-the-art trace-driven emulation tool called Saturator \cite{180329}, originally designed for cellular networks, works for Wi-Fi, and provide suggestions on how such a tool can be adapted to correctly record Wi-Fi traces. The rest of this paper is organized as follows: In Section~\ref{sec:related}, we highlight the existing work. Section~\ref{sec:background} explains the challenges in using existing methods. In Section~\ref{sec:exp-setup}, we explain our measurement set-up. Section~\ref{sec:sat-over-wifi} presents the results of the state-of-the-art trace-driven emulation tool \cite{180329} over Wi-Fi. In Section~\ref{sec:conclusion}, we conclude our work and mention future work.

\section{Related Work}
\label{sec:related}


\paragraph{\textbf{\textit{Simulation}}} There are several network simulators. NS-2 \cite{ns2} and NS-3 \cite{ns3} are open-source network simulators to reproduce Internet systems. OMNET++ \cite{omnet++} is a simulation platform for building simulators for wireless, wired, and queuing networks amongst others. OPNET \cite{opnet} is an open-source network simulation tool that offers various topologies and configurations. NETSIM \cite{netsim} is a commercial network simulator that provides simulation for layer 1 and layer 2 capabilities of Wireless Local Area Network (WLAN). QualNet \cite{Qualnet:2014} is a commercial network simulator for scalable network technologies. It offers a Graphic User Interface (GUI) to make things easier for users as there is no coding involved. TraceReplay is an application layer simulator built in NS-3 for network traces \cite{7561126}. Despite the fact that there are a lot of simulators available, it is always very hard to get realistic settings.

\paragraph{\textbf{\textit{Testbeds}}} Wireless Hybrid Network (WHYNET) \cite{Zhou:2006:WHT:1160987.1161016} is a hybrid testbed as it allows the use of simulation, emulation, and real hardware. It allows the integration of these on both individual and combined levels. There is limited remote access to the WHYNET testbed infrastructure. ORBIT \cite{1424763} is a radio grid emulator which provides the functionality of reproducing wireless experiments with a large number of nodes. It allows the introduction of fading and controlled interference. MONROE \cite{8406836} is an open-access hardware-based measurement platform for doing experiments on mobile broadband. The advantages of testbeds include running experiments over real wireless links and remote management. However, the testbeds come with a few drawbacks like no repeatability, no mobility (of nodes), small-level scaling, and dependency on location.

\paragraph{\textbf{\textit{Emulation}}} There are several network emulators that have been previously used to emulate network conditions for Wi-Fi and other technologies. Mobile network tracing \cite{Noble:1996:MNT:RFC2041} observes traffic passively to generate traces and then uses Packet Modulator (PaM) to corrupt, delay or drop captured packets. However, mobile network tracing does not address the question of different machines sharing the same bandwidth. Trace-modulation \cite{Noble:1997:TMN:263109.263140} listens to a path passively multiple times to generate traces of real network behavior. Common Open Research Emulator (CORE) \cite{4753614} is a network emulator that boasts a GUI that helps in drawing topologies. While CORE emulates layer 3 and above, Extendable Mobile Ad-hoc Network Emulator (EMANE) \cite{5379781} emulates physical and data link layers (1 and 2). Mahimahi\cite{Netravali:2015:MAR:2813767.2813798} is a framework for recording and replaying HTTP traffic under different network conditions. Mahimahi uses DelayShell and LinkShell for emulating a fixed propagation delay, and fixed and variable capacity links respectively. MpShell \cite{Deng:2014:WLB:2663716.2663727} extends the Mahimahi framework to record Wi-Fi and LTE traces simultaneously. This work was mainly developed to evaluate the performance of MP-TCP in different network conditions and for various types of applications. A mobility emulation framework called EmuWNet \cite{8407025} is proposed for signal propagation measurements in wireless networks. It allows users to replay measurement traces collected either by simulations or real-world experiments. It is based on ORBIT \cite{1424763} testbed and offers various mobility scenarios for testing in a controlled environment.

In this paper, we opt for trace-driven simulation and the state-of-the-art trace-driven emulation tool, called Saturator \cite{180329}. Trace-driven emulation is a good option because 1) testing takes place on a real network and 2) traces help in repeatability; the results and testing environment can be reproduced later. We, therefore, prefer trace-driven emulation to simulation, testbeds, and experimentation.

\section{Background}
\label{sec:background}
\hfill\

In cellular networks, the only form of congestion at the base-station is self-induced congestion. Further, in cellular networks, the uplink and downlink communications of users take place on different time slices and they do not interfere with each other. There are rarely any standing queues created by the traffic of other users in the cell. Even in the case of individual queues, a queuing delay of 750 ms does not starve the load \cite{180329}. Whereas the medium is shared in Wi-Fi and hence, the queues at every network entity are shared by all users. Cellular networks are also more robust because there are several numbers of retransmissions to cope with packet loss, which is not the case with Wi-Fi. The throughput can only be affected by the demand and competition for allocation of the time slices in cellular networks, whereas there are other factors that affect throughput in Wi-Fi (including cross traffic). This introduces unique challenges for recording Wi-Fi traces.


\begin{table}[!b]
	\begin{subtable}{.4\linewidth}
		\centering
		\begin{tabular}{cc}
			\toprule Network (s) & Traces (s) \\
			\midrule 0.40 & 0.40\\
			0.29 & 0.29\\
			0.32 & 0.32\\
			0.37 & 0.37\\
			0.37 & 0.37\\
			\bottomrule
		\end{tabular}
		\caption{LTE}
	\end{subtable}
	\begin{subtable}{.7\linewidth}
		\centering
		\begin{tabular}{cc}
			\toprule Network (s) & Traces (s) \\
			\midrule 0.24 & 0.32\\
			0.18 & 0.31\\
			0.20 & 0.50\\
			0.24 & 0.30\\
			0.22 & 0.32\\
			\bottomrule
		\end{tabular}
		\caption{Wi-Fi}
	\end{subtable}
	\caption{File transfer completion times in seconds}
	\medskip
	\small
	This table shows the transfer completion time for downloading a file over LTE and Wi-Fi (Network columns) versus over-recorded and replayed LTE and Wi-Fi traces (Traces columns).
\end{table}

We use the state-of-the-art tool Saturator \cite{180329} to demonstrate its behavior over cellular network and Wi-Fi. Saturator consists of two sender programs running at a client and a server. The client is connected via two cellphones, one cellphone is used to saturate the uplink and the downlink, while the second cellphone is used for feedback. A window of N packets is maintained by each sender program. Using the feedback packets, each sender adjusts the window size to ensure that the link is saturated without causing any self-induced packet loss. Both the client and the server record the timestamp, sequence number, and estimated round-trip time (RTT) or one-way delay for every received data or ACK packet in their respective logs. Uplink and downlink latency, throughput, and packet loss can be computed using these logs. The feedback in Saturator consists of ACK packets sent to the sender for the packets received by the receiver. The sender can then keep sending consistently to saturate the link reliably. Therefore, a separate interface is needed for feedback to ensure the timely delivery of ACK packets to the sender, and to avoid any impact of feedback delay on the link saturation. If the interface that has to be saturated, is also used for feedback, queuing might cause enough delay for ACK packets to not arrive on time. In this case, there is a possibility the link might not get properly saturated.

The traces collected via Saturator are replayed using Cellsim \cite{180329}. Cellsim requires its own machine with two Ethernet interfaces. It is connected to the client directly with an Ethernet cable, whereas it is connected to the Internet with a second Ethernet interface. The client machine is not connected to the Internet, whereas the server machine is connected to the Internet via Ethernet. Cellsim delays the packets received on both its Ethernet interfaces by a considerable amount of time to emulate propagation delay before adding the packets to the queue. The traffic from the client is sent to the server over the Internet by Cellsim.

\begin{figure}[t!]
	\centering
	\includegraphics[scale=0.2]{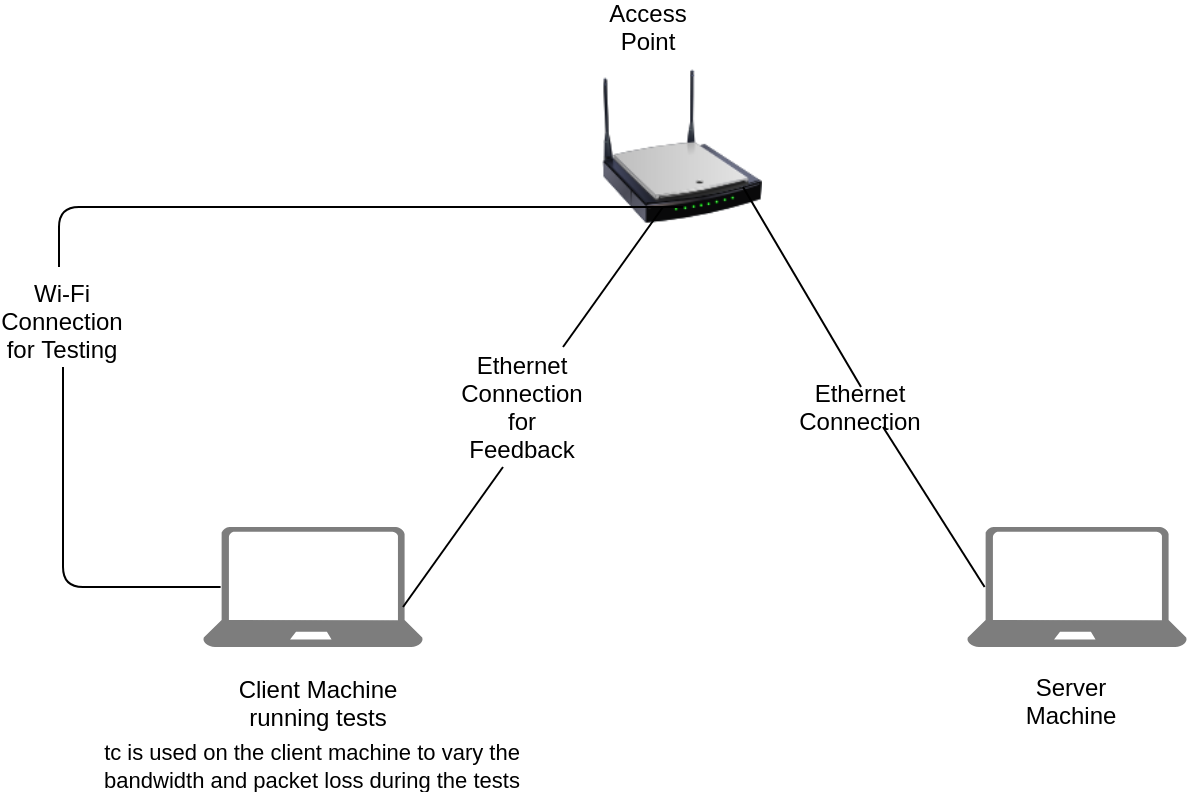}
	\caption{Experiment Set-up}
	\medskip
	\small
	This figure shows the experimental set-up for our tests that includes 2 Dell laptops and a TP-Link access point
\end{figure}

\begin{figure}[b!]
	\centering
	\includegraphics[scale=0.5]{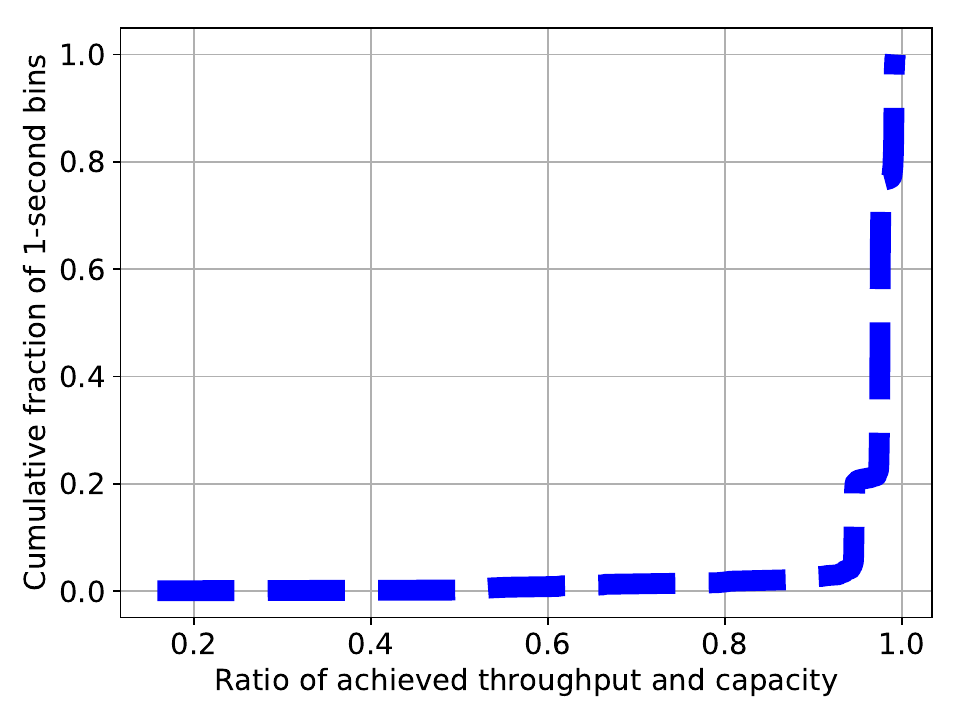}
	\caption{Distribution of the fraction of capacity consumed by Saturator}
\end{figure}

We collect the traces with Saturator for both LTE and Wi-Fi. We then use Cellsim to replay the traces captured with Saturator. We set up 3 machines for replay; one as a client, one as a server, and one as a Cellsim machine. We measure the time it takes to download a given file over a Wi-Fi network and compare the results against the time it takes to download the exact same file using traces recorded with Saturator for the same Wi-Fi network. We repeat the same measurement for LTE. To ensure that the LTE and Wi-Fi conditions do not fluctuate much between the real and trace-driven experiments, we run the real and emulated experiments back-to-back. We use a 250 MB file for the Wi-Fi experiment, and a 15 MB file for LTE, given that we have limited data for LTE.


We see in Table 1 that the file transfer completion times in both record and replay are always exactly the same for LTE. This is expected because Saturator is designed for cellular networks. Whereas, for Wi-Fi, it takes more time to complete the transfer in the replay phase. This indicates that Saturator works really well for LTE as expected but it is most probably not measuring the Wi-Fi link properly. There is likely to be some error in the Wi-Fi measurements that creates doubt about Saturator's compatibility with Wi-Fi.


\section{Experimental Set-up}
\label{sec:exp-setup}
\hfill\

\begin{figure}[b!]
	\centering
	\includegraphics[scale=0.44]{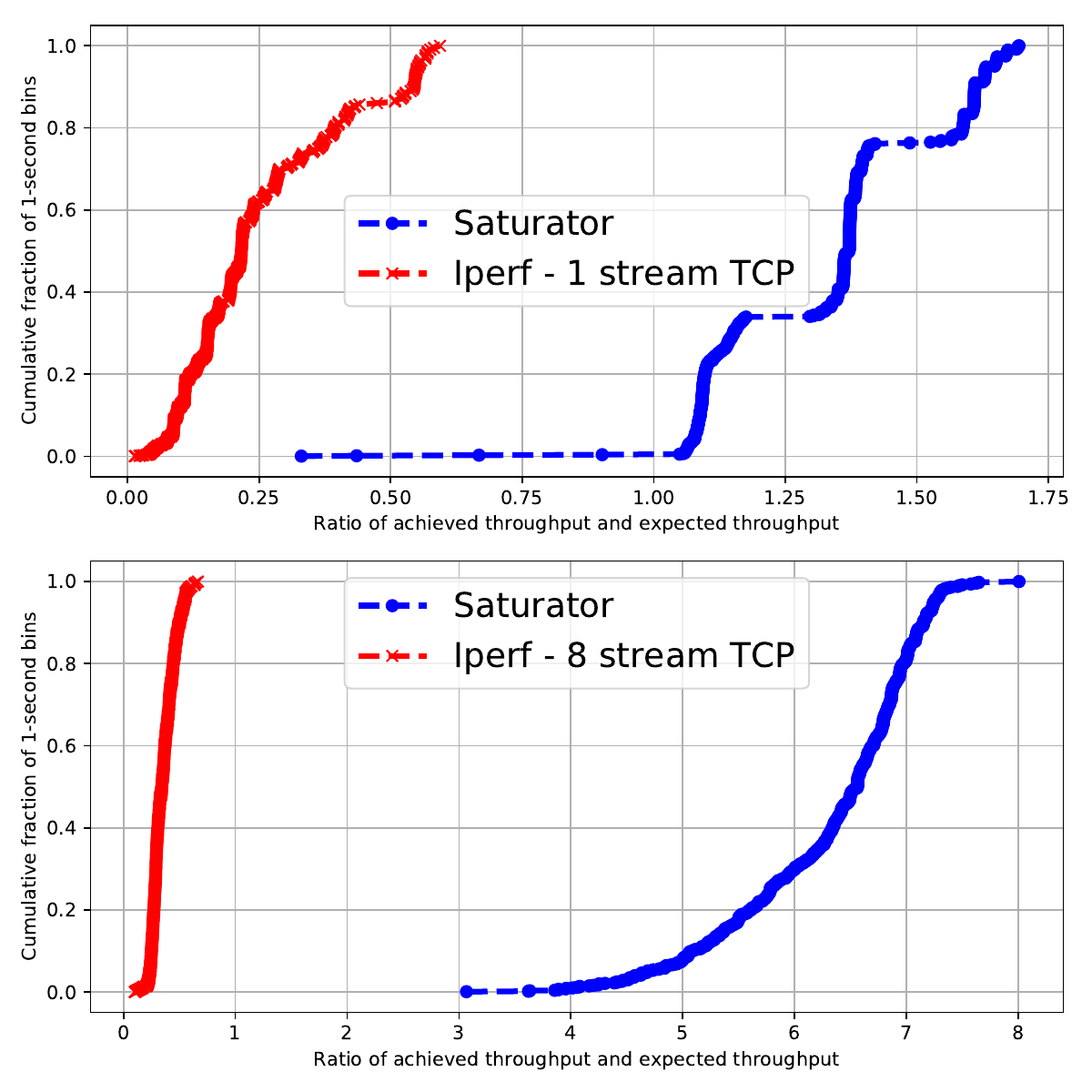}
	\caption{Saturator with concurrent TCP traffic and variable bandwidth}
\end{figure}

In this section, we present the experimental set-up that we use for all our measurements. Figure 1 presents our measurement set-up which is as follows:
\begin{itemize}[noitemsep,nolistsep]
	\item one laptop as a client; connected to Wi-Fi for measurements and connected to an access point with an Ethernet cable for feedback.
	\item one laptop as server; connected to the same access point as the client with an Ethernet cable
\end{itemize}

\bigskip
The AP is TP-Link Archer C7 which supports 802.11ac, and 2.4 and 5 GHz connections simultaneously. We perform our testing with 802.11ac and 2.4 GHz. We use 2 Dell laptops with identical specifications; Intel 8th generation core i7 CPU - 1.9GHz (Turbo 4GHz), 16 GB RAM, and 520 GB SSD hard drive. Both machines have Ubuntu 18.04 freshly installed, they do not have anything else installed on them. We use this set-up to avoid any impact of CPU load on Saturator.

We run Saturator client and server versions on these machines. We use Linux's \textit{traffic control (tc)}~\cite{tc} option on the client machine to vary the bandwidth and loss rate. We need to limit the bandwidth to a certain value to test Saturator's compatibility with Wi-Fi. We vary the bandwidth values every 12 seconds as follows: 15 Mbps, 40 Mbps, 10 Mbps, 30 Mbps, and 15 Mbps.

Similarly, for loss rate, we use \textit{tc} and vary the percentage value for packet loss every 12 seconds as follows: 0.3\%, 0.5\%, 0.25\%, 1\%, and 0.3\%.

We perform all the tests 5 times each once we have all set-up in place. The test duration is 1 minute for all the tests. We conduct tests for constant and variable bandwidth and packet loss values.

\section{Saturator over Wi-Fi}
\label{sec:sat-over-wifi}
\hfill\


\begin{figure}[b!]
	\centering
	\includegraphics[scale=0.44]{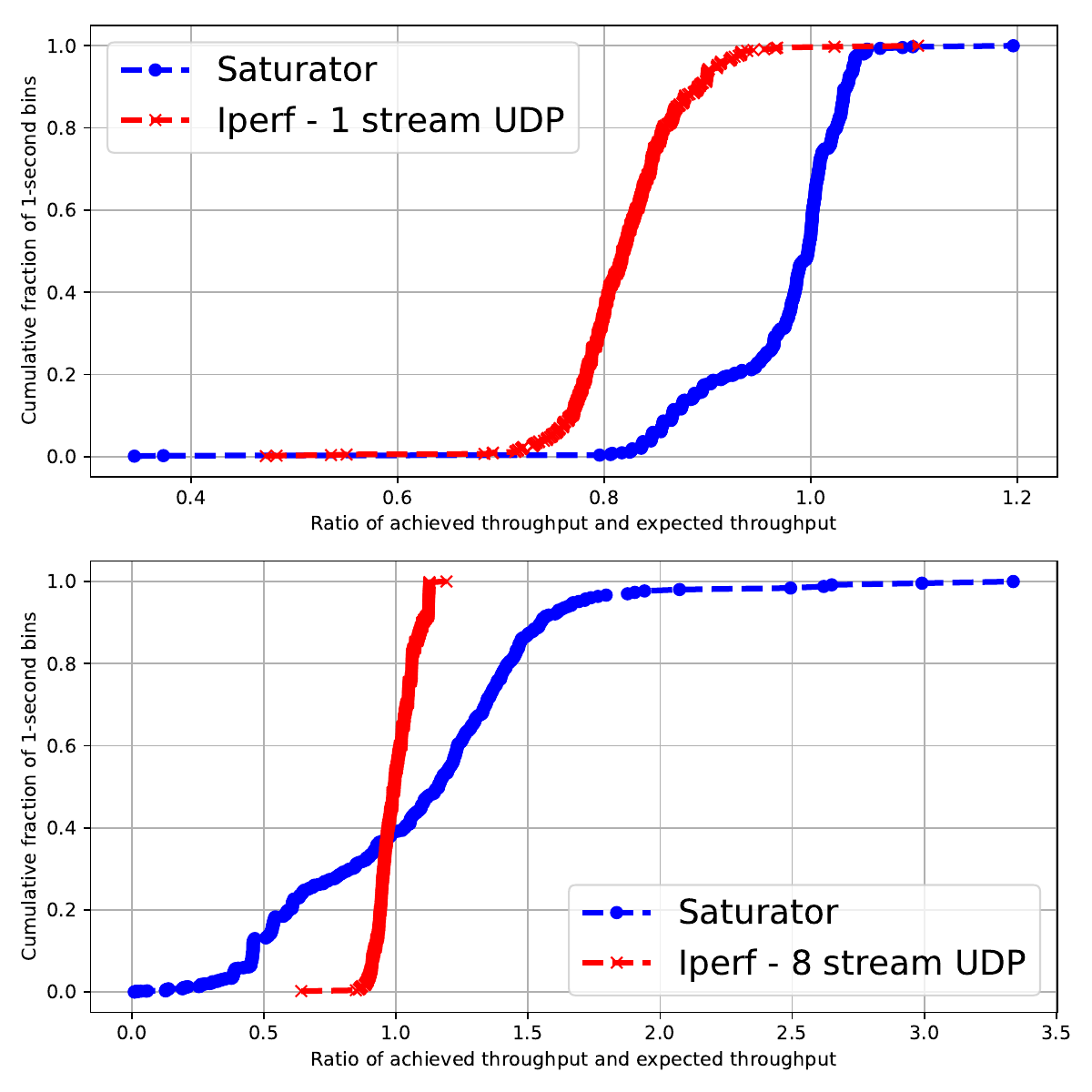}
	\caption{Saturator with concurrent UDP traffic and variable bandwidth}
	\medskip
	\small
	The effect of Saturator on concurrent UDP traffic in the presence of variable bandwidth is represented in this figure
\end{figure}

\begin{figure}[t!]
	\centering
	\includegraphics[scale=0.4]{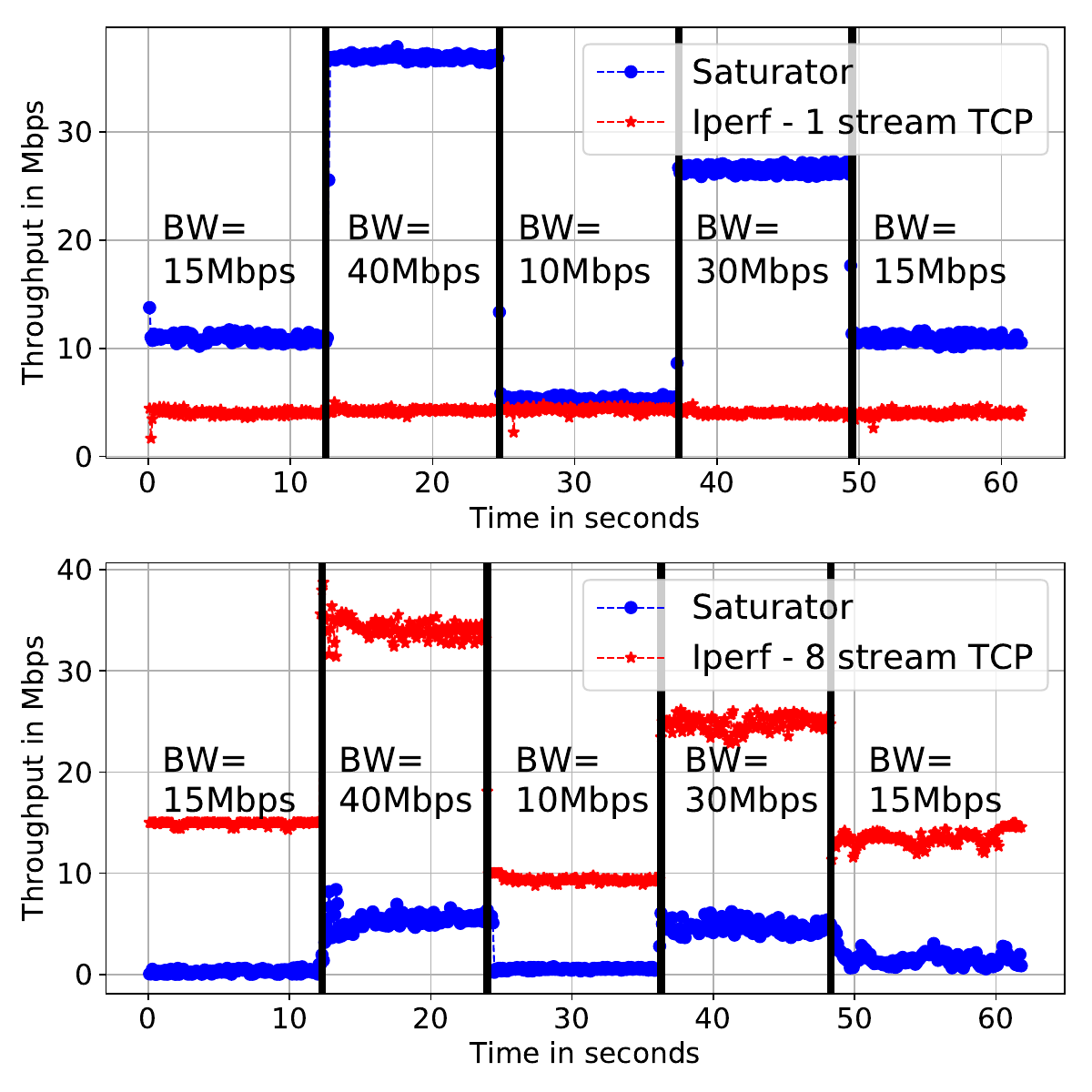}
	\caption{Saturator with concurrent UDP traffic and variable bandwidth}
	\medskip
	\small
	BW means bandwidth and it represents the bandwidth limit set by \textit{tc}
\end{figure}

In this section, we showcase the limitations of Saturator for Wi-Fi. Given that Wi-Fi's queuing mechanism and medium access control are different from that of Cellular, we study the impact of these two features on Saturator's ability to accurately record Wi-Fi traces. First, we evaluate Saturator's ability to saturate the Wi-Fi link without any concurrent traffic, a set-up that closely resembles Cellular links. Next, we evaluate the behavior of Saturator when we introduce concurrent UDP and TCP traffic on the Wi-Fi link. Finally, we look at the impact of saturating both the Wi-Fi uplink and downlink simultaneously.

\subsection{Saturator with Concurrent Traffic}
\label{sec:concurrent-traffic}
\hfill\

The first step is to test Saturator with Wi-Fi without any modifications and in the absence of cross-traffic; to verify if Saturator is able to saturate the link. We make use of \textit{tc} to limit the bandwidth as explained in the previous section and carry out tests without any concurrent traffic.

\begin{figure}[t!]
	\centering
	\includegraphics[scale=0.4]{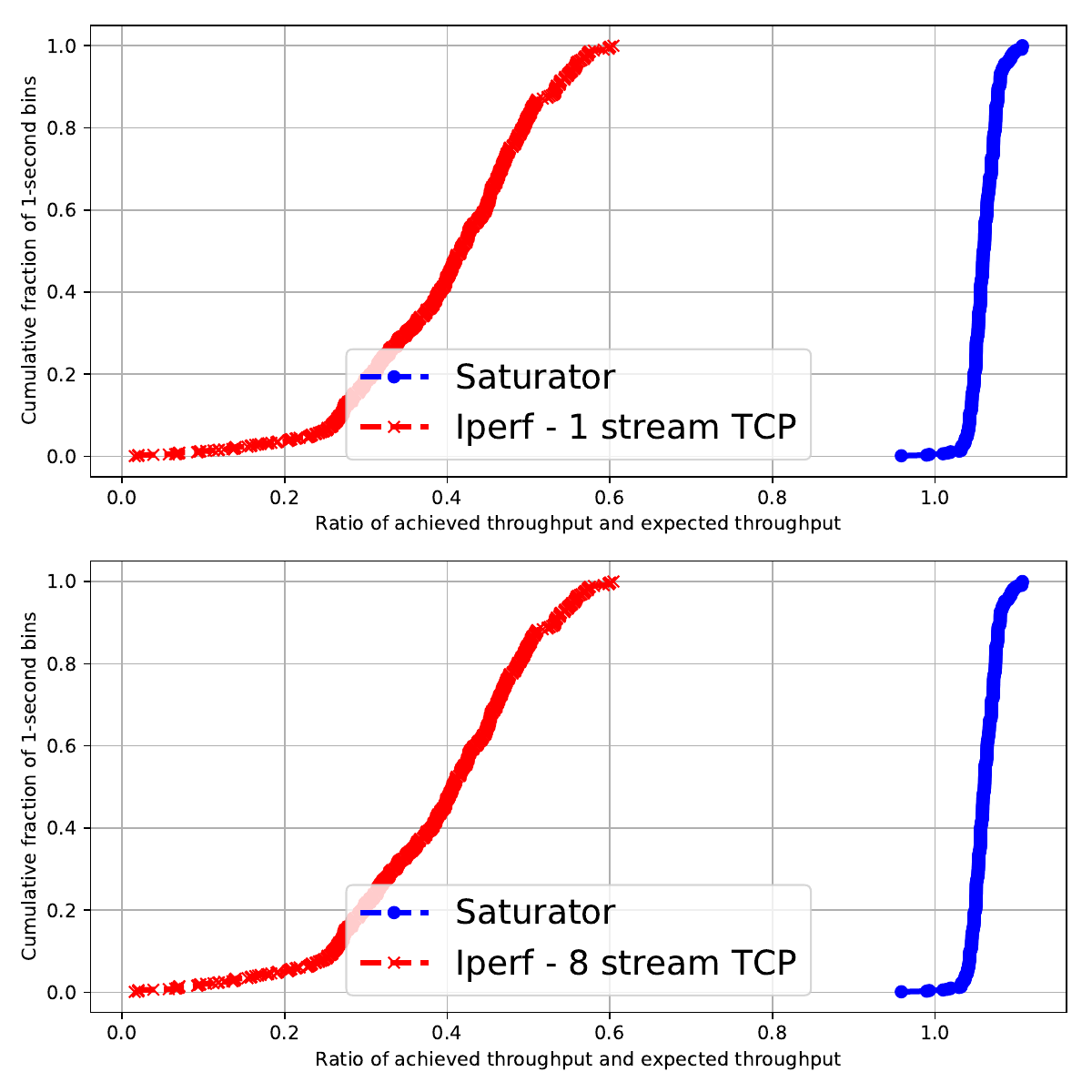}
	\caption{Saturator with concurrent TCP traffic and variable packet loss}
\end{figure}

We observe that Saturator is able to fill the pipe. Saturator reacts to the bandwidth variations very well and adapts accordingly. This is the expected behavior because the testing conditions without concurrent traffic are similar to cellular networks. Figure 2 shows the ratio of achieved throughput and capacity. The more the ratio is closer to 1, the more Saturator is able to fill the pipe. Smaller values of the ratio in Figure 2 can be deceptive because Saturator is able to consistently saturate the link. These small values of the ratio, however, are the result of changing bandwidth during the tests as Saturator takes nearly 1–2 milliseconds to adapt to the new value.

The main question is how would Saturator cope with concurrent traffic in Wi-Fi. We consider the following conditions while doing these tests:
\begin{itemize}[noitemsep,nolistsep]
	\item with concurrent traffic and variable bandwidth
	\item with concurrent traffic and variable packet loss
\end{itemize}

\bigskip
We generate concurrent TCP and UDP traffic with iPerf~\cite{iperf}, and we limit the per-iPerf stream bandwidth to 5 Mbps. We see in Figure 3 that the ratio between achieved throughput and expected throughput is mostly close to 1 for Saturator; it does considerably well to fill the pipe with concurrent TCP traffic even with variable bandwidth. However, it ends up suppressing everything else, as TCP traffic gets a lot less than what is expected. As evident from Figure 3, the ratio for iPerf is way less than 1. iPerf gets around 1 Mbps for 1 stream, whereas it gets a maximum of 10 Mbps for 8 streams. It shows Saturator is not fair to TCP traffic.

The results are, however, different for concurrent UDP traffic generated by iPerf. When we use just 1 stream restricted to 5 Mbps, iPerf consistently manages to achieve the expected throughput. However, as we increase the number of streams for UDP traffic, iPerf's UDP traffic seems to saturate the pipe completely, as we observe in Figure 4. However, we see in Figure 4 that the ratio of achieved throughput and expected throughput for Saturator even exceeds 1; which suggests Saturator achieves the expected throughput. Figure 5 clears this anomaly; we see that Saturator is able to achieve the expected throughput for higher bandwidth values (i.e. 40 Mbps and 30 Mbps). However, iPerf UDP dominates for lower values of bandwidth (i.e. 15 Mbps and 10 Mbps). This raises a question of Saturator's compatibility with Wi-Fi. Figure 5 shows the result of one of the experiments but the results are consistent across all experiments. As discussed earlier, concurrent traffic is very likely to be present and Wi-Fi is not a stable medium as well, it casts doubt on the use of Saturator.

\begin{figure}[t!]
	\centering
	\includegraphics[scale=0.4]{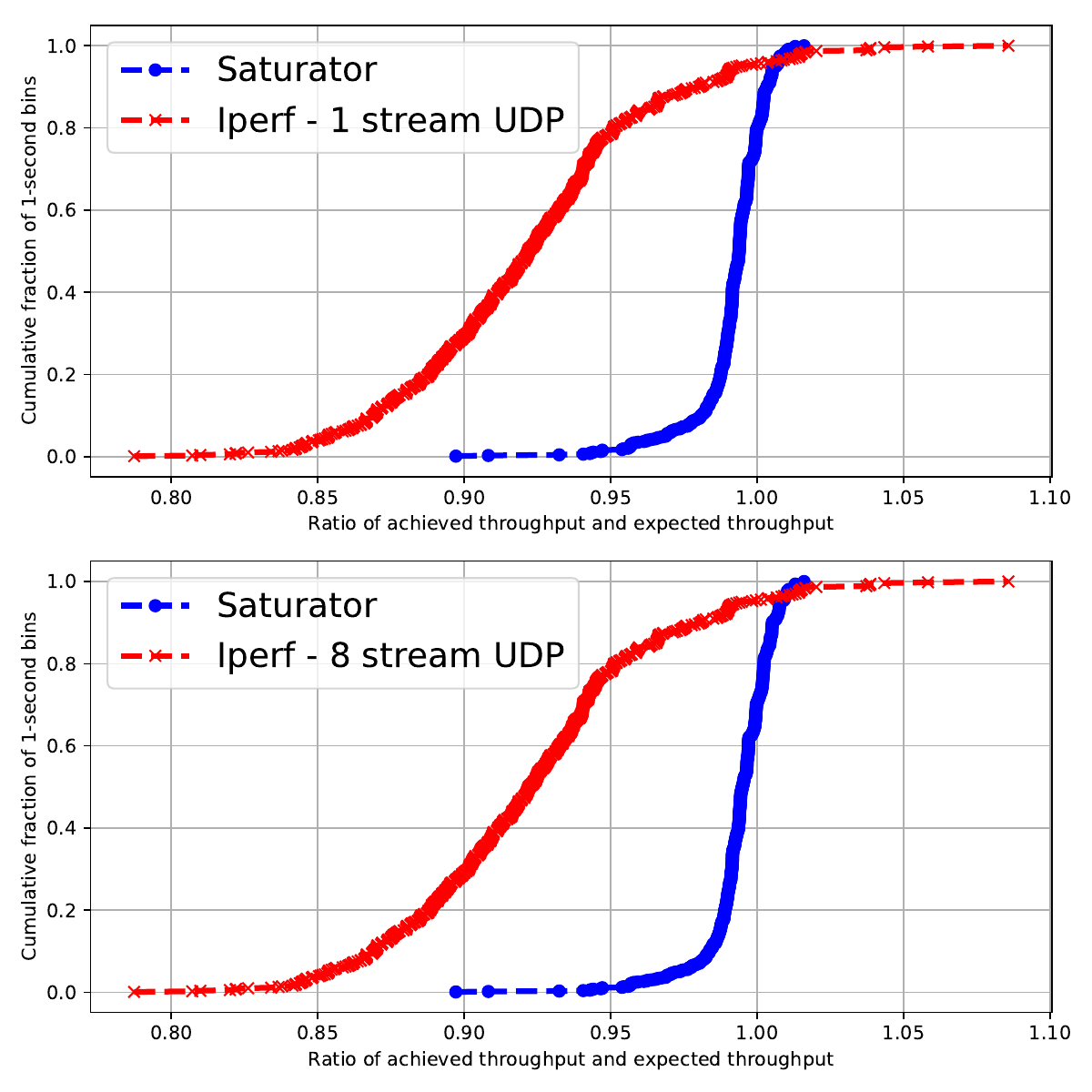}
	\caption{Saturator with concurrent UDP traffic and variable packet loss}
\end{figure}

Another important aspect to study about Saturator is its ability to react to packet loss. We use \textit{tc} to introduce packet loss as explained in Section~\ref{sec:exp-setup}.

\begin{figure}[b!]
	\centering
	\includegraphics[scale=0.5]{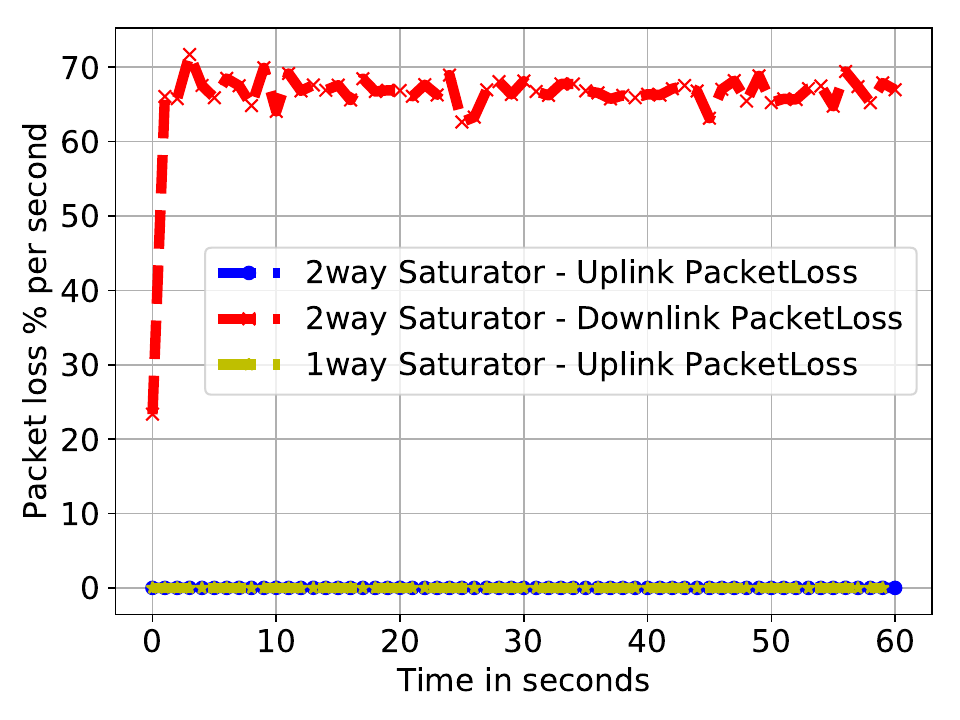}
	\caption{Saturator Packet Loss}
	\medskip
	\small
	This figure represents the packet loss calculated from Saturator logs
\end{figure}

\begin{figure}[b!]
	\centering
	\includegraphics[scale=0.5]{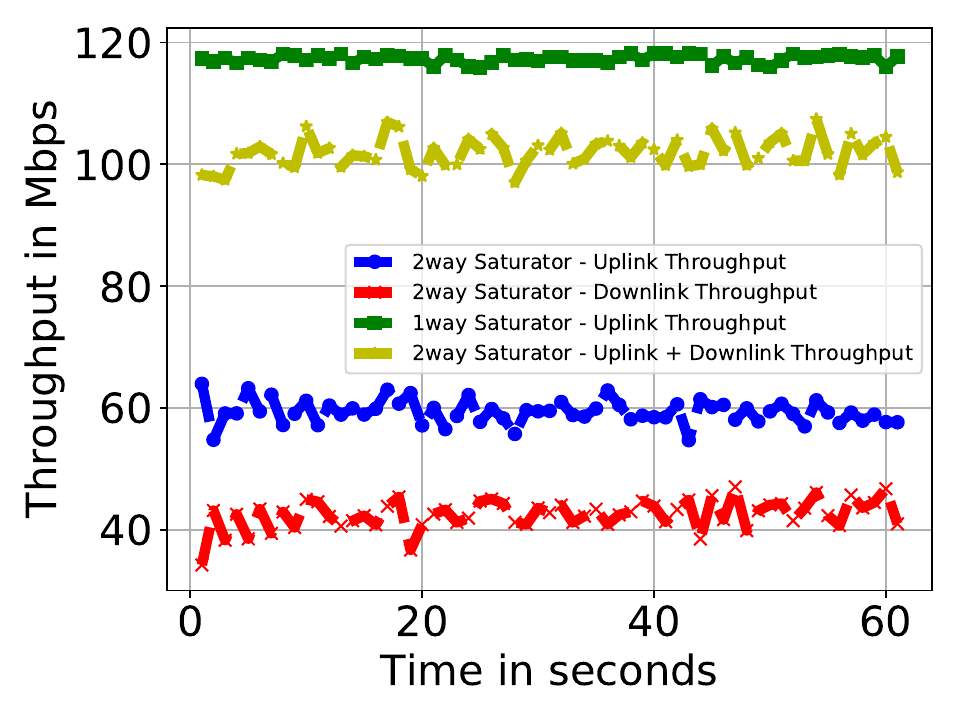}
	\caption{Saturator Throughput}
	\medskip
	\small
	This figure shows the comparison of 2-way Saturator throughput versus 1-way Saturator throughput
\end{figure}

We observe that Saturator is able to cope with the packet loss reasonably well as compared to the concurrent traffic. It is evident from Figures 6 and 7 that it keeps sending more traffic to reach a decent throughput, as the ratio between achieved throughput and expected throughput is consistently close to 1. It is a surprising finding considering 1) it is UDP traffic and 2) Saturator does not react to packet loss itself. We examine it further to find the cause of this behavior and analyze the logs. As mentioned in Section~\ref{sec:background}, Saturator reacts to RTT only, it just keeps the check of the number of packets in flight with respect to the window size and keeps sending packets whenever there is an opportunity. As packets get lost, there are fewer packets in flight and Saturator sees it as an opportunity to send more. In this way, it ends up sending more packets than it normally does in case of no loss. The sender side has a number of packets sent more than the number of packets received at the receiver by a certain percentage in accordance with the loss.

\subsection{Saturator in One Direction}
\label{sec:one-direction}
\hfill\

As mentioned earlier, Saturator is originally designed to work with cellular network and it saturates both uplink and downlink at the same time. It does not matter much for cellular networks because both communications take place on different time and frequency slots and they do not affect each other. Whereas Wi-Fi is a shared medium and there are always chances that uplink and downlink communications happening at the same time can interfere with each other resulting in decreased performance. We, therefore, make a minor change in Saturator to make it work in only one direction. We evaluate Saturator with Wi-Fi for 2 conditions as following:
\begin{itemize}[noitemsep,nolistsep]
	\item saturate both uplink and downlink
	\item saturate just uplink
\end{itemize}

\begin{figure}[t!]
	\centering
	\includegraphics[scale=0.5]{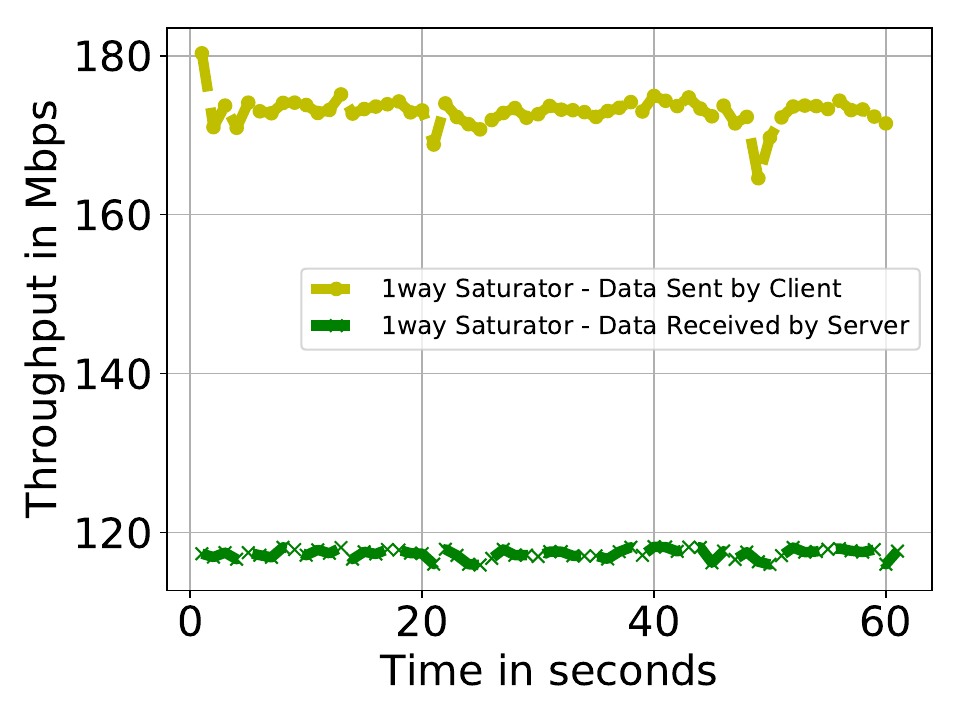}
	\caption{Data sent vs data received - 2-way Saturator}
\end{figure}

\bigskip
In the two-way experiments where Saturator is run in both uplink and downlink directions, it always gets more throughput on the uplink. Downlink communication is badly affected by packet loss as packet loss goes up to 80\%. Figure 8 represents the packet loss for one of the five experiments, it is consistent across all five runs. Uplink packet loss for the 2-way Saturator is not clearly visible in the figure because it is exactly the same as the 1-way uplink packet loss. The client always initiates the communication, which could explain why the uplink achieves more bandwidth than the downlink. We make a slight change to make sure the client just initiates the connection with the server but it actually starts sending data with a delay of 1 second. We find out that the server takes over the bandwidth initially but as soon as the client starts sending, we see more traffic on the uplink, same as in previous cases. Multiple factors could explain this behavior; one possible explanation is that the uplink queue is larger than the downlink queue. Another possible reason is that the access point might be giving higher priority to uplink traffic than downlink traffic.

\begin{figure}[t!]
	\centering
	\includegraphics[scale=0.53]{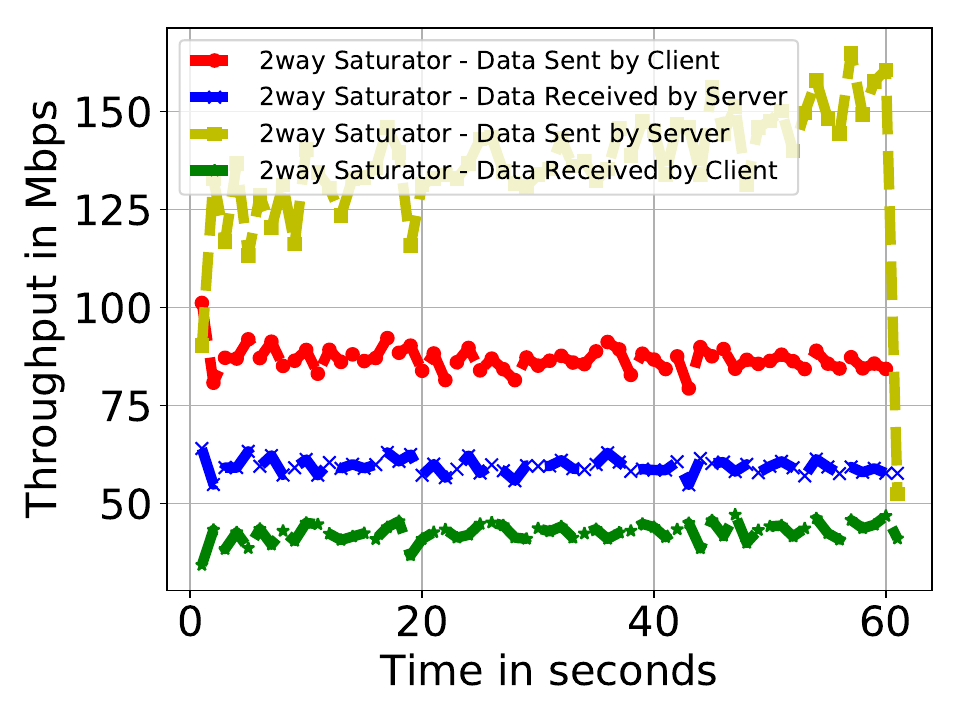}
	\caption{Data sent vs data received - 1-way Saturator}
\end{figure}

\begin{figure}[b!]
	\centering
	\begin{subfigure}[b]{0.4\textwidth}
		\centering
		\includegraphics[width=\textwidth]{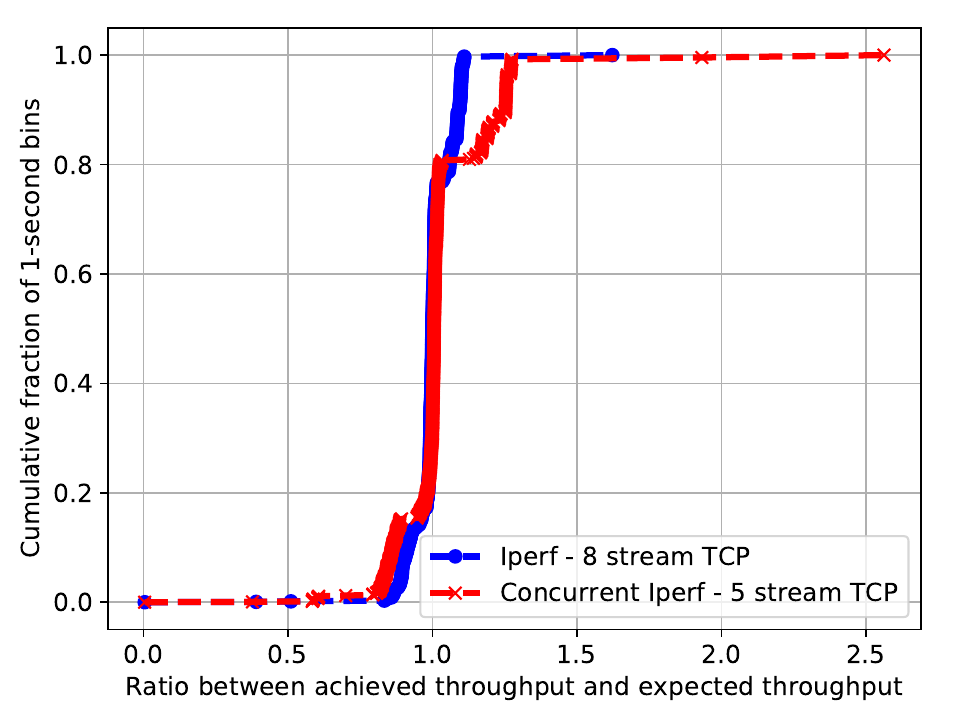}
		\caption{5 stream TCP Traffic}
	\end{subfigure}
	\begin{subfigure}[b]{0.4\textwidth}
		\centering
		\includegraphics[width=\textwidth]{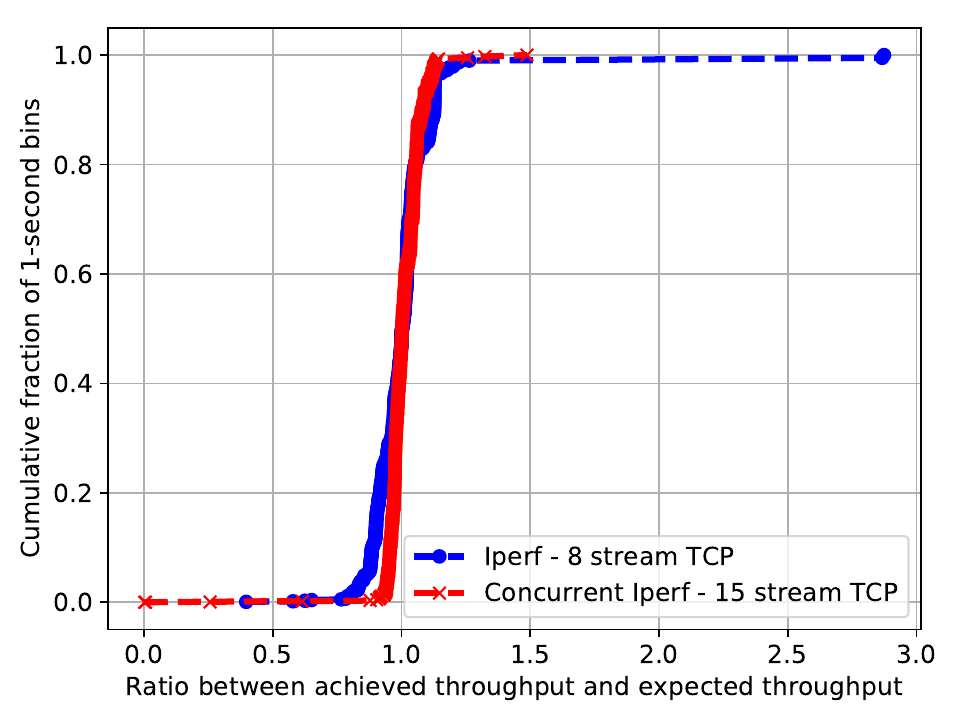}
		\caption{15 stream TCP Traffic}
	\end{subfigure}
	\caption{Multithreaded TCP iPerf - TCP vs TCP}
\end{figure}

Figure 9 shows the throughputs in Mbps for one run. The results are similar for all runs. We see that the results are far better and more stable when only the uplink is saturated. The 1-way throughput is always more than the sum of uplink and downlink throughputs of the 2-way Saturator. We recommend using Saturator only in the uplink direction when used with Wi-Fi to eliminate the possible interferences and collisions between the uplink and downlink traffic and enable Saturator to fully saturate the pipe.

\subsection{Packet Loss Issue}
\label{sec:packet-loss}
\hfill\

As we discussed earlier, there is a huge packet loss with Saturator running in both directions. We use the sequence numbers of packets received to find the packet loss. We are still not fully convinced that Wi-Fi could introduce such losses; so, we analyze it further to find out if it is really a Wi-Fi-related behavior. Since Saturator only logs incoming data packets at the Sender side, we use \textit{tcpdump}~\cite{tcpdump} to be able to capture the client's outgoing data traffic too. We use \textit{tshark}~\cite{tshark} to get timestamps and packet lengths from the pcap files and then use those to find the throughput. We can see in Figures 10 and 11 that Saturator sends way more than what is actually received. As mentioned previously, Saturator is designed for cellular networks, it does not consider Wi-Fi network conditions and ends up sending more than what it could actually send. This results in the driver dropping the packets to cope with excessive Saturator traffic. Therefore, the packet loss we see is not because of Wi-Fi, but it is because Saturator's sending window is tuned for LTE buffer sizes, which are typically larger than Wi-Fi's.

\begin{figure}[t!]
	\centering
	\begin{subfigure}[b]{0.4\textwidth}
		\centering
		\includegraphics[width=\textwidth]{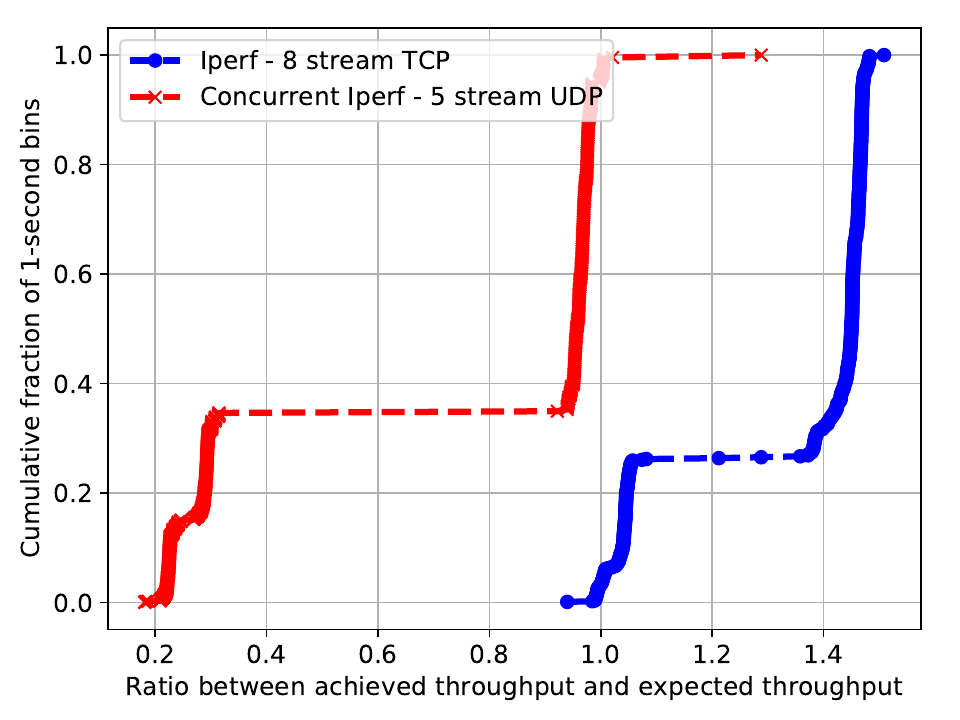}
		\caption{5 stream UDP Traffic}
	\end{subfigure}
	\begin{subfigure}[b]{0.4\textwidth}
		\centering
		\includegraphics[width=\textwidth]{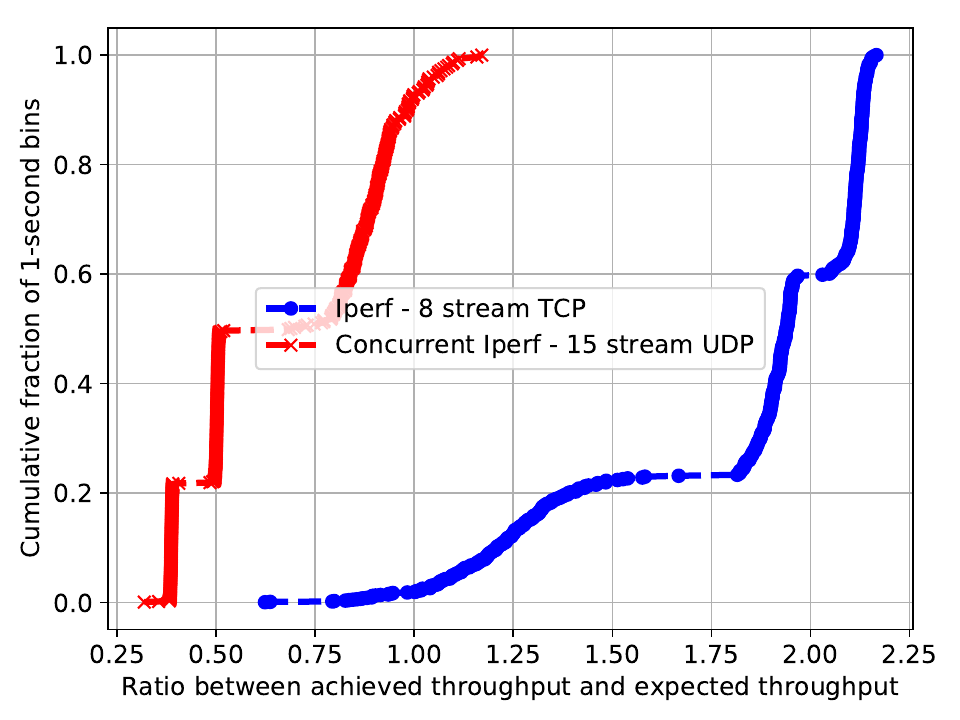}
		\caption{15 stream UDP Traffic}
	\end{subfigure}
	\caption{Multithreaded TCP iPerf - TCP vs UDP}
\end{figure}

\subsection{Alternative Solutions}
\label{sec:alternative}
\hfill\

Given the aforementioned limitations of Saturator, in this section, we explore whether iPerf could be used instead to record Wi-Fi traces. We study the impact of the number of threads of TCP iPerf on the bandwidth it measures. We use 8 iPerf TCP streams as traffic generated by us, and a different number of streams for TCP and UDP iPerf as concurrent traffic.  We consider the following scenarios:
\begin{itemize}[noitemsep,nolistsep]
	\item 8 iPerf TCP streams vs 5 iPerf TCP streams
	\item 8 iPerf TCP streams vs 15 iPerf TCP streams
	\item 8 iPerf TCP streams vs 5 iPerf UDP streams
	\item 8 iPerf TCP streams vs 15 iPerf UDP streams
\end{itemize}

\bigskip
We run \textit{tcpdump} in parallel with iPerf and use the pcap files to calculate the throughput. We see the results for all these scenarios in Figures 12 and 13. The bandwidth measured by iPerf depends on the number of streams used. If iPerf has more streams than concurrent traffic, then it ends up getting more bandwidth (mainly because of TCP's fair share); that would affect the real traffic in the network and create a bias in the measurements. It shows one has to be careful when choosing the number of parallel streams as it can not only create a bias in the measurements but also have a negative effect on the actual traffic; especially if multiple streams are used with UDP iPerf, it can badly degrade performance for other users.

\section{Conclusion}
\label{sec:conclusion}
\hfill\

In this paper, we evaluate how well the state-of-the-art trace-driven emulation tool Saturator captures Wi-Fi traces. Since Saturator is originally designed for cellular networks, we showcase how the differences between Wi-Fi and Cellular inhibit the applicability of Saturator for Wi-Fi as-is. We highlight through experimental analysis that the measurements done by Saturator are influenced by the nature of concurrent traffic. Additionally, we illustrate how saturating Wi-Fi in just one direction eliminates interference, thereby enhancing Saturator's ability to effectively fill the pipe.

Although all the experiments we conduct are in ideal Wi-Fi conditions, for future work, we aim to repeat our experiments in different set-ups to study the effects of multipath fading. We also plan to study the impact of 802.11 frame-aggregation on trace replay. Further, we plan to explore the idea of introducing packet loss in the replay as this is a common behavior in Wi-Fi. Finally, we aim to investigate the ideal window size to eliminate the observed packet drops at the driver side.

\Urlmuskip=0mu plus 1mu\relax
\bibliographystyle{acm}
\bibliography{INRIAThesis}

\end{document}